# An Impedance control approch of Ultra High Power Electric Arc Furnace

Yun Chol Guk, Sin Yong Chol, Kwak Son Il*

***Kim Il Sung** University*

*Ryongnam －Dong, Taesong District, Pyongyang, Democratic People's Republic of Korea*

*Email Address: ryongnam18@yahoo.com*

**Abstract** An approach is proposed to design the intelligent electrode position controller for UHP by using nonlinear scaling and fuzzy self－tuning PID control algorithm.

First, nonlinear scaling of controlled variable that compensate the nonlinearity of the object is proposed. Second, a fuzzy self－tuning PID electrode position control algorithm is designed and the parameters of the fuzzy inference are optimized by using GA (Genetic Algorithm). Finally, the effectiveness of the proposed approach is verified by field test.

**Keywords** : steel, electric arc furnace, UHP, electrode position control, fuzzy PID control, process control

## 1. Introduction

Ultra High Power Electric Arc Furnace (UHP) has become one of main means for the steel industry since they have high productivity, low cost of production and high quality of products. To enhance the quality of electrode position control is an essential issue in raising the productivity, lowering the cost of production and safe operation of UHP.

From the viewpoint of the control theory, UHP can be seen a time-variant, nonlinear control object.

Various modeling methods [1-6] are proposed for modeling electric arc furnace; for example, adaptive arc furnace model [1], adaptive neuro-fuzzy inference system [2] and chaos-based model [3].

And for electrode position control, PI control [8], model-based predictive control [9], and various intelligent control approaches (for example, fuzzy logic control [10], neural network control [11] and variable-universe fuzzy control [12]) are proposed.

The mode of electrode position control according to the controlled variable can be classified into current settling mode[11], impedance settling mode[4], power settling mode[8]. The impedance settling mode control is frequently used because it can solve the problem of "unbalanced load". But in the case of impedance settling mode control, the nonlinear relation between the arc current and the impedance must be considered.

In this paper, 1st, we propose a nonlinear scaling that can compensate the nonlinear relation between the arc current and the impedance. 2nd, we design a fuzzy self-tuning PID controller for electrode position control. Finally, we verify the effectiveness of the proposed approach through the field test for a UHP.

## 2. The electrode position control system

**a) Recognition of melting stages**

Generally, the melting process of arc furnace is divided into melting－down stage, oxidation stage and reduction stage, and, according to the melting stages, the characteristics of arc furnace is changed in wide range[7]. Thus we estimate the melting stages by using experience equation automatically and use it in the control algorithm.

The factors for recognition of melting process are integrated active power, power-on time, power-off time, amount and ratio of raw material, etc. And among them, dominant factors are integrated active power and power-off time.

From a viewpoint of the energy flow in furnace, some of throwing electric energy is lost through joule heat of the resistance of transformer and lead and some of thermal energy that is produced due to the resistance of electric arc is lost through thermal radiation and thermal convection. Only the remaining energy is effectively used for melting scrap and heating up of the melted iron. This is simply expressed as Eq (1).

$$W_{active} = \eta W_{throw} - cT_{off} \tag{1}$$

Where, $W_{active}$ [kWh] is effectively used thermal energy, $W_{throw}$ [kWh] is integrated active power, $\eta$ is efficiency factor of electric energy, $T_{off}$ [min] is power-off time and $c$ [kWh/min] is loss of thermal energy per 1 minute of power-off time.

The required thermal energy that used for melting scrap at melting−down stage is denoted by $W_0$ [kWh] and a new variable (as an alias "process variable") $p$ is defined as

$$p = \frac{\eta}{W_0} W_{throw} - \frac{c}{W_0} T_{off} \tag{2}$$

Then an estimating equation for melting−down stage is obtained as Eq (3).

$$\begin{aligned} p < 1 &\Rightarrow Melting-down\ period \\ p \geq 1 &\Rightarrow Melting-down\ period\ is\ over \end{aligned} \tag{3}$$

We obtained experience equation for estimating melting−down stage as Eq (4) by using operating data of dozens of charges and System Identification Toolbox of MATLAB.

$$p = 5.162 \times 10^{-5} W_{throw}(t) - 8.641 \times 10^{-4} T_{off}(t) \tag{4}$$

Like above, an estimating equation for oxidation stage and reduction stage can be obtained (Eq (5), (6)).

$$q = 9.423 \times 10^{-5} W_{throw}(t) - 1.949 \times 10^{-3} T_{off}(t) \tag{5}$$

$$\begin{aligned} q < 1 &\Rightarrow before\ Reducing\ period \\ q \geq 1 &\Rightarrow Reducing\ period \end{aligned} \tag{6}$$

**b) Nonlinear scaling of the error of impedance**

The electrode position control are classified to current settling mode, impedance settling mode and electric power settling mode according to the selected controlled variable.

The current settling mode stabilize only the arc currents, but does not consider the phase voltages. So the voltages and electric powers of 3 phases are not harmonized. That is, "the problem of unbalanced 3 phases" is occurred. But the impedance settling mode stabilize the impedence (the ratio of phase voltage to arc current), therefore, it can harmonize not only the arc currents of 3 phases but also phase voltages and electric powers of 3 phases.

Thus the impedance settling mode has an advantage that can solve "the problem of unbalanced 3 phases", and it has been used frequently in electrode position control.

But the impedance settling mode has some problems because there exists an nonlinear relation between the arc current and the impedance as follow Equation (4).

That is, when the value of impedance has been settled at the neighborhood of the setpoint, the average value of the arc current has become bigger than the setpoint of the arc current corresponding to the setpoint of the impedance (Fig 5).

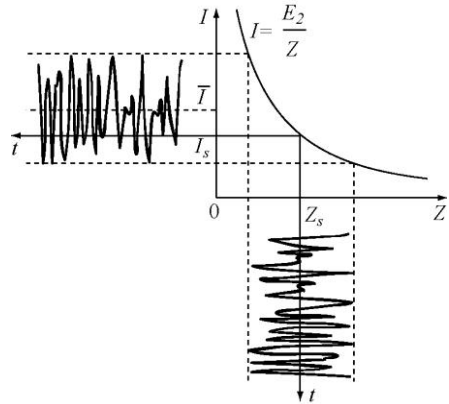

Figure 5  Nonlinear relation between impedance and arc current

(When the value of impedance has been settled at the neighborhood of the setpoint, the average value of the arc current has become bigger than the setpoint of the arc current corresponding to the setpoint of the impedance)

This phenomenon can be seen more clearly at the melting－down stage than the reduction stage and the oxidation stage, because the fluctuation of the arc current is more excessive during the melting－down stage. In the previous works, this problem had been solved by choosing the setpoint of the impedance somewhat bigger than theoretically calculated value. But the fluctuation of the arc current varies constantly, so the static error of the arc current could not be eliminated in this way.

To solve this problem, we do not use the impedance $Z$ as the controlled variable, but we propose a new controlled variable $y(Z)$ that satisfies the Eq (7).

$$\begin{cases} \dfrac{dy(Z)}{dI} = c \ (const) \\ y(\infty) = 0 \\ y(Z_s) = Z_s \end{cases} \quad (7)$$

The new controlled variable $y(Z)$ that satisfies Eq (7) is calculated as follows.

$$y(Z) = \frac{Z_s^2}{E_2} I = \frac{Z_s^2}{Z} \quad (8)$$

The error of the new controlled variable from setpoint is expressed as the Eq (9).

$$e = y - y_s = \frac{Z_s}{Z}(Z_s - Z) \quad (9)$$

From the Eq (9), we can see that the error of the new variable is a nonlinearly scaled error of the impedance. In other words, it can be compensated that the nonlinear relation between arc current and impedance by assigning a smaller weight for impedance bigger than the setpoint, and, a bigger weight for impedance smaller than the setpoint.

If we design a stabilizing control which leads the nonlinearly scaled impedance to zero, it results in the linear relation between the new controlled variable $y$ and arc current $I$ as Eq (8) and, in turn, the average value of arc current is settled at the neighborhood of its set point (Fig 6).

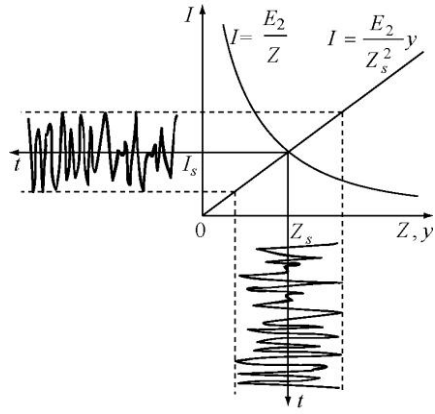

Figure 6. The linear relation between the new controlled variable and arc current

(The average value of arc current is settled at the neighborhood of it's setpoint during impedance is settled at the neighborhood of its set point)

In the next subsection, we propose a fuzzy self－tuning PID control algorithm that stabilize the new control vartiable at the setpoint.

### c) Fuzzy self－tuning PID control algorithm

From the viewpoint of the control theory, the electrode position control object can be seen the time－varying, nonlinear system. That is, the effect of the arc length to the electric circuit is constantly varied during the whole melting process (Eq (2)), and the electric circuit involves a nonlinearity (Eq (3), (5)).

Considering this, here we propose a fuzzy self－tuning PID control algorithm for electrode position control. PID controller performs the electrode position control and the parameters of the PID controller are adjusted by fuzzy inference, according to the melting process and control error.

First, to avoid serious vibration and large overshoot caused by excessive integral action, we use the real differential PID control algorithm with separated integral action (Eq (10)).

$$u(s) = \begin{cases} K_P \left( 1 + \dfrac{1}{T_i s} + \dfrac{T_d s}{1 + \dfrac{T_d}{n} s} \right) e(s), & |e| \leq \delta \\ K_P \left( 1 + \dfrac{T_d s}{1 + \dfrac{T_d}{n} s} \right) e(s), & |e| > \delta \end{cases} \qquad (10)$$

Where, $n$ is settled as 10 and $\delta$ is a boundary value for separation of the integral action. Actually we choose $n = 10$, $\delta = 0.1 y_s$.

Practically, we use the difference equation (11)－(14).

$$u(k) = \begin{cases} u_p(k) + u_i(k) + u_d(k), & |e(k)| \leq \delta \\ u_p(k) + u_d(k), & |e(k)| > \delta \end{cases} \qquad (11)$$

$$u_p(k) = K_P e(k) \qquad (12)$$

$$u_i(k) = K_I T_c \sum_{i=1}^{k} e(i) \qquad (13)$$

$$u_d(k) = \lambda u_d(k-1) + (1-\lambda)K_D \frac{(e(k) - e(k-1))}{T_c} \quad (14)$$

Where, $K_I = \frac{K_P}{T_i}$, $K_D = K_P \cdot T_d$, $\lambda = \frac{T_d}{T_c n} / \left(1 + \frac{T_d}{T_c n}\right)$ and $T_c$ is control cycle.

Next, PID parameters like as $K_P$, $K_I$, $K_D$ are adjusted constantly by fuzzy inference. The fuzzy inference rules for tuning the PID parameters are as Eq (15).

$$CR_l : \text{If } p \text{ is } P_i \text{ and } y \text{ is } Y_j$$
$$\text{Then } K_P = K_P^l, K_I = K_I^l, K_D = K_D^l, \quad l = 1, 2, \cdots, 15 \quad (15)$$

Where, $p$ is the early defined "process variable" (that is gradually larger from 0 when melting process started and become 1 when oxidizing stage.), $y$ is nonlinearly scaled impedance and $K_P^l, T_I^l, T_D^l$ are singletons of the PID parameters of $l^{th}$ fuzzy inference rule.

The fuzzy sets of the premise variables are as Fig 7 and the boundaries of the fuzzy sets are as table 2 and table 3.

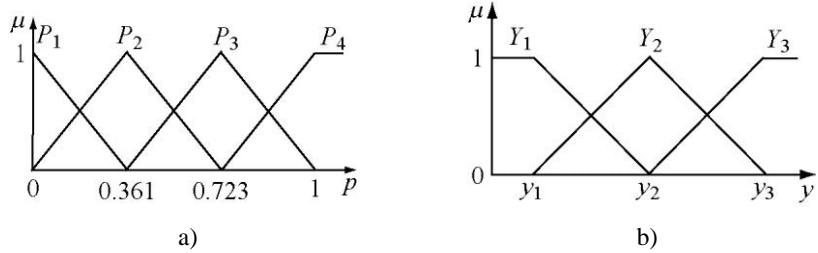

Fig 7. Fuzzy sets of the premise variables

a) Fuzzy sets of process variable $p$          b) fuzzy sets of control variable $y$

Table 2. The relation between $p$, $\beta$ and melting stages

| $p$ | $P_1$ | $P_2$ | $P_3$ | $P_4$ | $P_5$ |
|---|---|---|---|---|---|
| $\beta$ | 12.0 | 9.0 | 6.0 | 3.7 | 1.2 |
| Melting stages | Melt−down stage | Melt−down stage | Melt−down stage | Oxidation stage | Reduction stage |

Table 3. the boundaries of the fuzzy sets of $y$

| boundaries / Melting stages | $y_1$ | $y_2$ | $y_3$ |
|---|---|---|---|
| Melt−down stage | 3.0 | 6.0 | 9.0 |
| Oxidation stage | 2.4 | 4.8 | 7.2 |
| Reduction stage | 2.0 | 4.0 | 6.0 |

Finally, PID parameters of controller are calculated by using Sugeno's reasoning power-off time as Eq (16).

$$K_P = \frac{\sum_{l=1}^{15} w_l K_P^l}{\sum_{l=1}^{15} w_l}, K_I = \frac{\sum_{l=1}^{15} w_l K_I^l}{\sum_{l=1}^{15} w_l}, K_D = \frac{\sum_{l=1}^{15} w_l K_D^l}{\sum_{l=1}^{15} w_l} \quad (16)$$

Where, $w_l = P_i(p) \cdot Y_j(y)$ is fitness of each rule, $P_i(p)$, $Y_j(y)$ are respectively membership degrees of $i^{th}$ fuzzy set of process variable $p$ and $j^{th}$ fuzzy set of control variable $y$.

The singletons for PID parameters of $l^{th}$ fuzzy rule $K_P^l, T_I^l, T_D^l$ are selected by using genetic algorithm (GA) so that ISE index is became optimal for local linear model of electrode lifting object.

Fuzzy inference rules for PID parameters $K_P^l, T_I^l, T_D^l$ are as tables 4, 5, 6.

Table 4. Fuzzy inference rules for $K_P$

| y \ p | $Y_1$ | $Y_2$ | $Y_3$ |
|---|---|---|---|
| $P_1$ | 0.3668 | 0.6655 | 1.140 |
| $P_2$ | 0.4879 | 0.8867 | 1.514 |
| $P_3$ | 0.7197 | 1.321 | 2.272 |
| $P_4$ | 1.517 | 3.016 | 7.025 |
| $P_5$ | 3.536 | 6.570 | 12.19 |

Table 5. Fuzzy inference rules for $K_I$

| y \ p | $Y_1$ | $Y_2$ | $Y_3$ |
|---|---|---|---|
| $P_1$ | 0.1067 | 0.0017 | 0.0036 |
| $P_2$ | 0.0081 | 0.0049 | 0.0043 |
| $P_3$ | 0.0020 | 0.0018 | 0.0044 |
| $P_4$ | 0.1030 | 0.0034 | 0.0115 |
| $P_5$ | 0.0186 | 0.0173 | 0.0070 |

Table 6. Fuzzy inference rules for $K_D$

| y \ p | $Y_1$ | $Y_2$ | $Y_3$ |
|---|---|---|---|
| $P_1$ | 0.0053 | 0.0010 | 0.0014 |
| $P_2$ | 0.0019 | 0.0010 | 0.0013 |
| $P_3$ | 0.0088 | 0.0030 | 0.0009 |
| $P_4$ | 0.0015 | 0.0017 | 0.0007 |
| $P_5$ | 0.0030 | 0.0029 | 0.0031 |

**d) Block diagram and control algorithm for the electrode position control system**

The block diagram for the impedance settling mode electrode position control based on fuzzy self－tuning PID control algorithm using nonlinear scaling is as Fig 8.

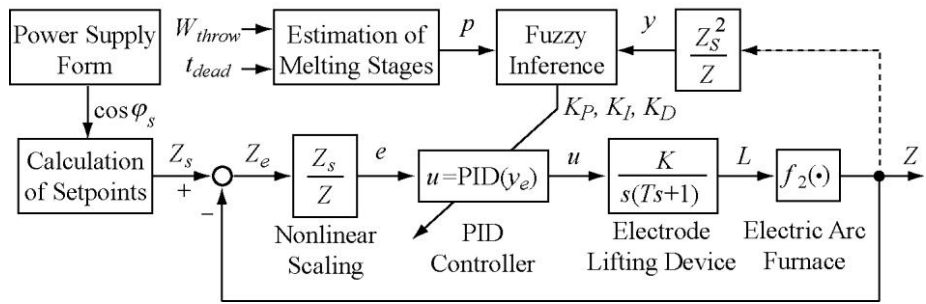

Fig 8. The block diagram for the electrode position control system

The calculation of set points block calculates the setpoint of impedance from the set value of power factor that presented at the power supply form. The estimation of melting stages block estimates the melting process by considering the integrated active power and power-off time. The PID controller calculates the driving voltages for the electrode lifting device by considering the nonlinearly scaled error of the impedance. The fuzzy inference block adjusts the parameters $K_P, T_I, T_D$ of the PID controller.

The algorithm of the impedance settling mode fuzzy self−tuning PID electrode position control is as follows.

Step 1. If not low voltage ($E_2(k) > 1.2IX$) and no arc ($I(k) < 5\,000\,A$), then lift down the electrode at high speed ($u(k) = -1.0V$).

Step 2. If over current ($I(k) > 1.25 \times I_S$), then lift up the electrode at a low speed ($u(k) = 0.3V$) until 1sec.

Step 3. If low voltage ($E_2(k) \leq 1.2I(k)X$) or danger current ($I(k) \geq 1.5 \times I_S$) or maintain over current ($I(k) > 1.25 \times I_S$) for 1sec, then lift up the electrode at high speed ($u(k) = 1.5V$).

If this situation continues during 2sec, then lift up all the electrodes at full speed ($u_1(k) = u_2(k) = u_3(k) = 2.5V$).

Step 4. If not low voltage ($E_2(k) > 1.2I(k)X$) and arc exists ($I(k) > 5000\,A$), then arc_success = 1.

Step 5. If arc_success == 1, then it is calculated the nonlinearly scaled error of the impedance and the accumulation value of the error (Eqs (27), (28)).

$$e(k) = \frac{Z_S}{Z(k)}(Z_S - Z(k)) \qquad (27)$$

$$S(k) = \sum_{i=1}^{k} e(i) \qquad (28)$$

Step 6. If $|e(k)| < 0.1Z_S$ and $|S(k)| < 100\,m\Omega$, then stop the electrode ($u(k) = 0V$). Otherwise calculate the control force by using the fuzzy self-tuning PID control algorithm ($u(k) = FPID(e(k))$).

Step 7. If end_control == 1 then finish the algorithm, or else go to the step 1.

In this algorithm, steps 1~3 are the part for prevent the electrode breaks, and steps 4~6 are the part for electrode position control.

The flow chart of this algorithm is as Fig 9.

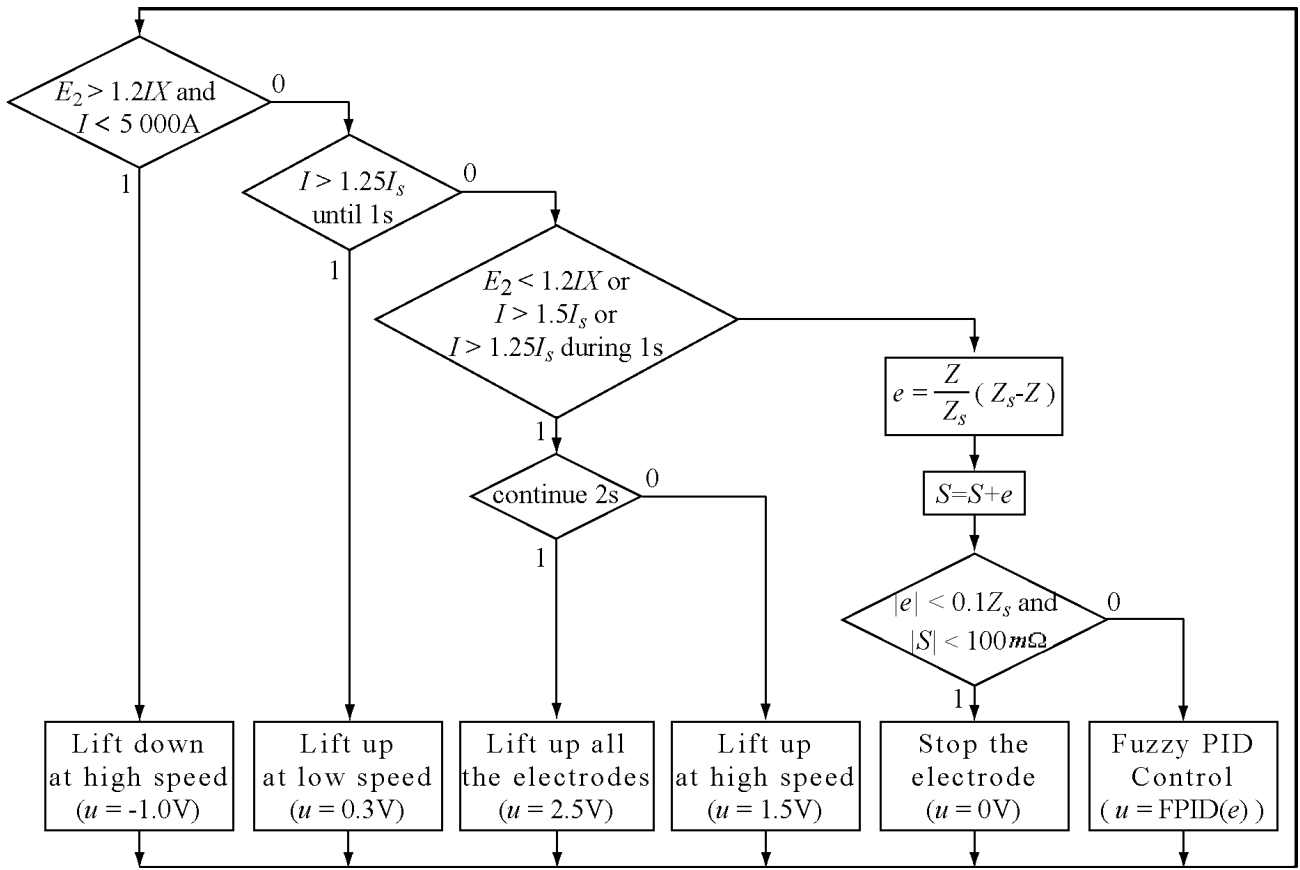

Fig 9. The flow chart of the electrode position control algorithm

## 3. Field test

The description of the UHP that we used for field test is as follows.

Capacity of the furnace 40t,

Output of molten iron 35t/charge,

Capacity of the transformer 26MVA,

Allowable current 30 000A,

Rated primary voltage 20KV,

Range of the secondary voltage 180~540V,

Number of taps of the transformer 17,

Number of taps of the reactor 7

The parameters of the electrode lifting device are as follows.

$$K = 15, \ T = 0.1$$

The effectiveness of the proposed method is compared with the methods (current settling mode PID control, impedance settling mode PID control). The result is as table 7. As you can see, the proposed method has small static error and reduce fluctuation of arc current remarkably.

Table 7. Comparision of cybernetical effectiveness between the previous/proposed method
A-current settling mode PID, B- impedance settling mode PID, C- proposed method

| evaluation criterion | | A | B | C |
|---|---|---|---|---|
| Static error of the arc current (A) | melting－down stage | －3 400 | 2 300 | 200 |
| | oxidation stage | －2 700 | 1 000 | 100 |
| | reduction stage | －600 | 300 | －100 |
| Standard deviation of the arc current (A) | melting－down stage | 4 200 | 4 500 | 3 400 |
| | oxidation stage | 3 600 | 3 900 | 3 200 |
| | reduction stage | 2 100 | 3 400 | 1 500 |

And the economical effectiveness of the proposed method is analyzed for dozens of charge. The result is that electric power consumption has been saved more than 30~50kWh/(steel 1t) and melting time has been saved more than 20~23 minutes.

## Conclusion

The mathematical model of the UHP was obtained and the setpoints of controlled variables wa calculated. A nonlinear scaling power-off time that compensates the nonlinear relation between the arc current and the impedance was proposed. The electric arc furnace is the time－varying, nonlinear system, so we designed the fuzzy self-tuning PID controller for electrode position control. The effectiveness of the proposed approach was verified through the field test for 40t UHP.